\documentclass[reprint,amsmath,amssymb,showpacs,aps,]{revtex4-1}
\usepackage{graphicx}
\usepackage{dcolumn}
\usepackage{bm}
\usepackage{soul}
\usepackage{ulem}
\usepackage{color}
\usepackage[T1]{fontenc}
\begin{document}
\title{Kinetically engendered sub-spinodal lengthscales in spontaneous dewetting of thin liquid films.}
\author{TirumalaRao Kotni}
\author{Jayati Sarkar}
\email{jayati@chemical.iitd.ac.in}
\author{Rajesh Khanna}
\email{rajkh@chemical.iitd.ac.in}
\affiliation{Department of Chemical Engineering, Indian Institute of Technology Delhi, New Delhi-110016}
\date{\today}

%%%%%%%%%%%%%%%%%%%%%%%%%%%%%%%%%%%%%%%%%%%%%%%%%%%%%%%%%%%%
\begin{abstract}

Numerical simulations of spontaneous dewetting of non-slipping, variable viscosity unstable thin liquid films on 
homogeneous substrates reveal the existence of sub-spinodal lengthscales through formation of satellite holes, a 
marker of nucleated dewetting and/or heterogeneous substrates, in the late stages of dewetting if the liquid 
viscosity decreases continually with decreasing film thickness.  
These films also show established signatures of slipping films such as faster rupture and flatter morphologies in the early 
stages  even without invoking any slippage.
\end{abstract}
%%%%%%%%%%%%%%%%%%%%%%%%%%%%%%%%%%%%%%%%%%%%%%%%%%%%%%%%%%%%
\pacs{68.15.+e, 47.20.Ma, 47.54.-r, 68.55.J-}
\maketitle

Spontaneous dewetting of supported thin liquid films 
\cite{Ruck1974,Brochard1990,Reiter1992,Khanna2000} 
is increasingly being harnessed to create template-free meso-nano patterned surfaces which find use in MEMS/NEMS, 
microfluidic devices, optoelectronic devices, anti-reflective coatings etc., 
\cite{Angew2013}.
As is the case, the lengthscales thus achieved are upper-bounded by a spinodal lengthscale which depends on the film 
thickness and liquid properties such as Hamaker Constants, hydrodynamic slippage at the solid interface, surface 
tension and mass density but is surprisingly independant of the elastic modulii
\cite{Kajari2004,Karin2005,Sharma2002,Sarkar2010}.
To achieve smaller lengthscales one has to necessarily rely on nucleated dewetting (through distribution of nucleating 
sites) 
\cite{Mecke2000} 
and/or heterogeneity of the underlying solid substrate (due to formation of satellite holes around already growing 
holes)
\cite{Sharma2000}. 
Thus, being able to spontaneously generate sub-spinodal lengthscales remains a persistent pursuit.
Elsewhere, thin liquid films experiments remain an invaluable probe to understand behavior of liquids in confinement
\cite{Gunter2013}.
Though a reconciliation of dewetting experiments with theory and simulations in terms of forces, length and time 
scales and morphology has already happened
\cite{Khanna1998,Hermighaus1998,Sharma2006},
recent reports about reduced viscosity in thinner polymer films 
\cite{Green2002,Bodiguel2006},
necessitate a recalibration.
The reason for this anisotropic viscosity as explained is twofold.
Primarily it is considered that near-to-surface polymer chains in coil-size thick layers shows accelerated reptation 
due to reduced contact with a very thin molten region between them and the substrate.
Secondly, exclusion of other chains due to reduced pervaded volume of the near-to-surface chains leads to lesser 
entanglement and allows for other modes of movements in addition to reptation thereby decreasing the viscosity near 
the substrate. 
The present letter principally aims to show that this decrease in viscosity with decreasing thickness leads to 
emergence of sub-spinodal lengthscales through formation of satellite holes without invoking heterogeneous substrate 
or nucleation.
Additionally, it shows that signatures of slipping films, faster rupture and flatter morphologies, are also present 
in the early stages of dewetting of these non-slipping films.
This makes interpretation of experiments with films such as polystyrene films even more complex as they are
likely to exhibit slippage as well as variable viscosity.
Though not expressely shown in the letter, due to constraints of prohibitively large simulations, the generation of 
multimodal distribution of lengthscales is a natural extension of its findings.

\begin{figure}[htbp]
\begin{center}
\begin{tabular}{cc}
\resizebox{2mm}{!}{A} &
\resizebox{2mm}{!}{B} \\
\resizebox{40mm}{!}{\includegraphics[width=1.6in]{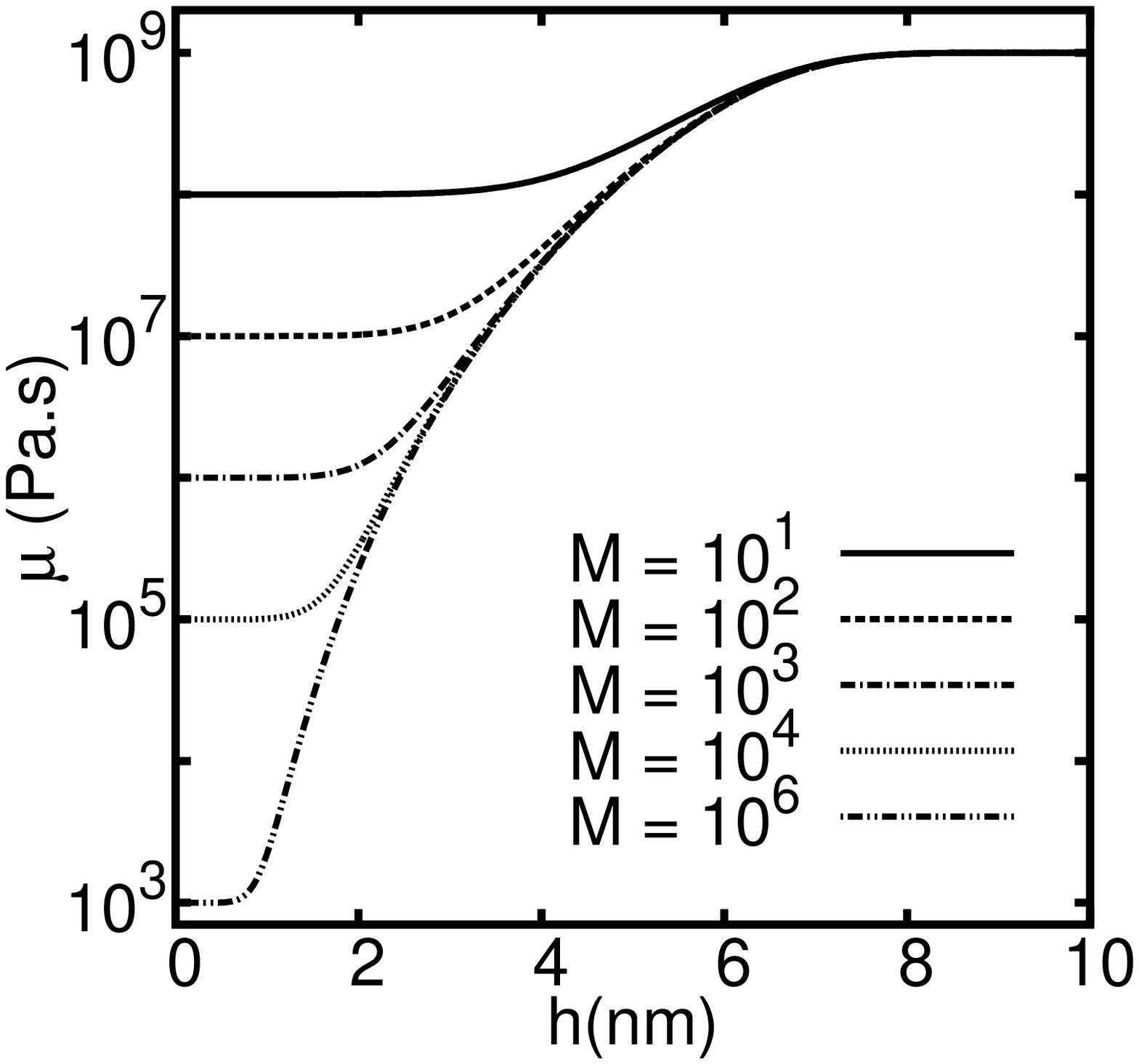}} &
\resizebox{40mm}{!}{\includegraphics[width=1.5in]{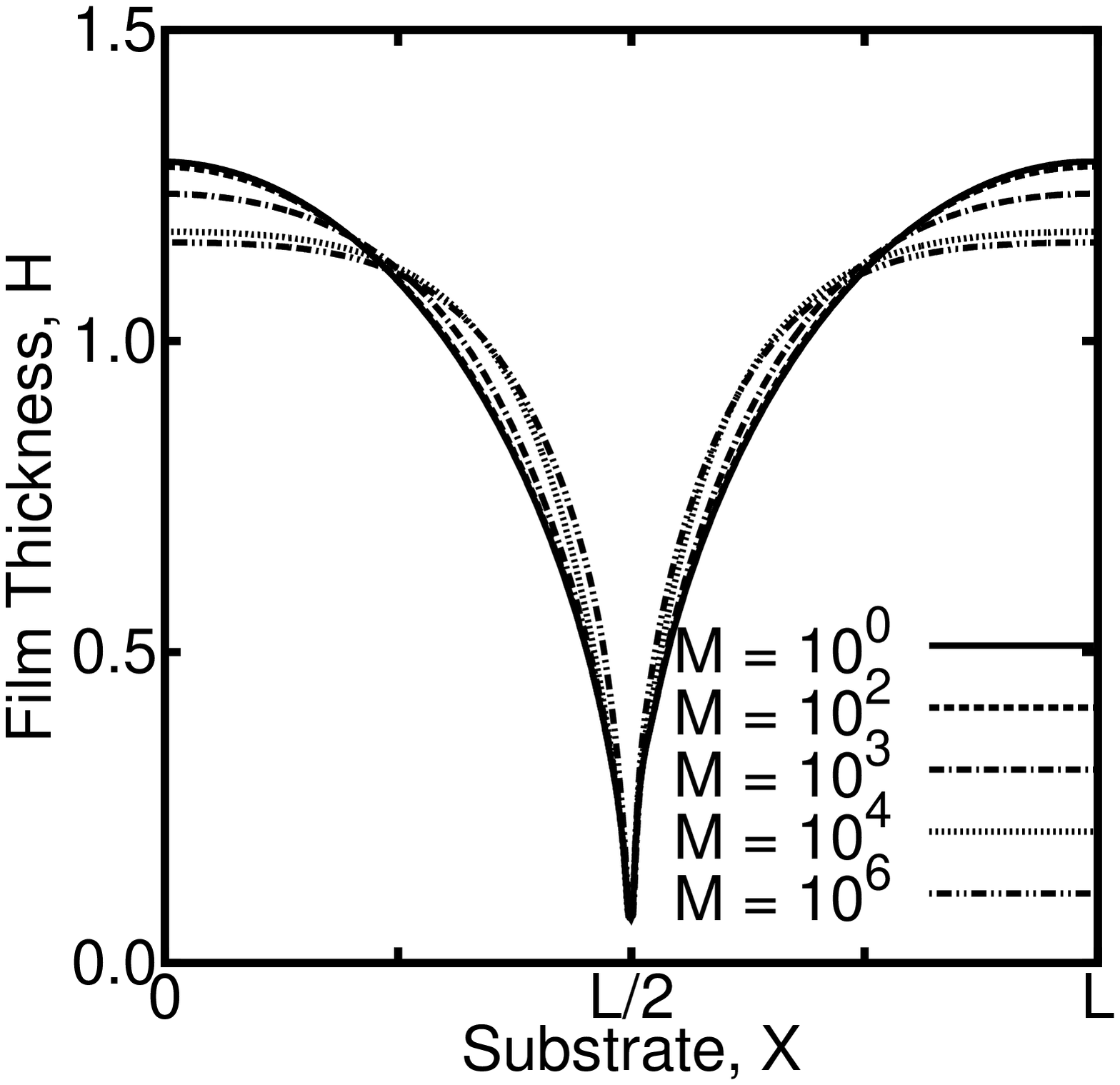}} \\
 & \\
\resizebox{2mm}{!}{C} &
\resizebox{2mm}{!}{D} \\
\resizebox{40mm}{!}{\includegraphics[width=1.6in]{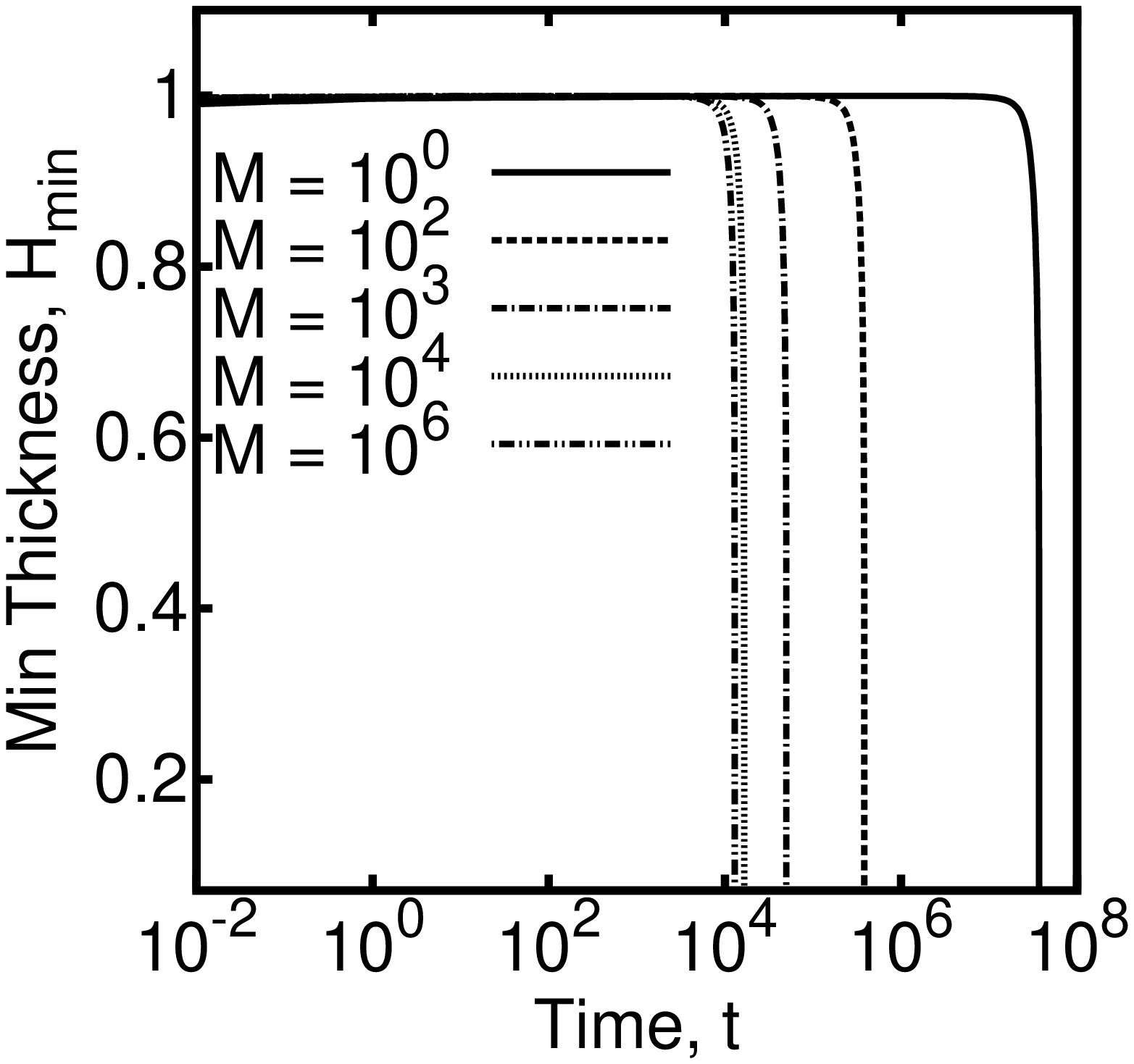}} &
\resizebox{40mm}{!}{\includegraphics[width=1.6in]{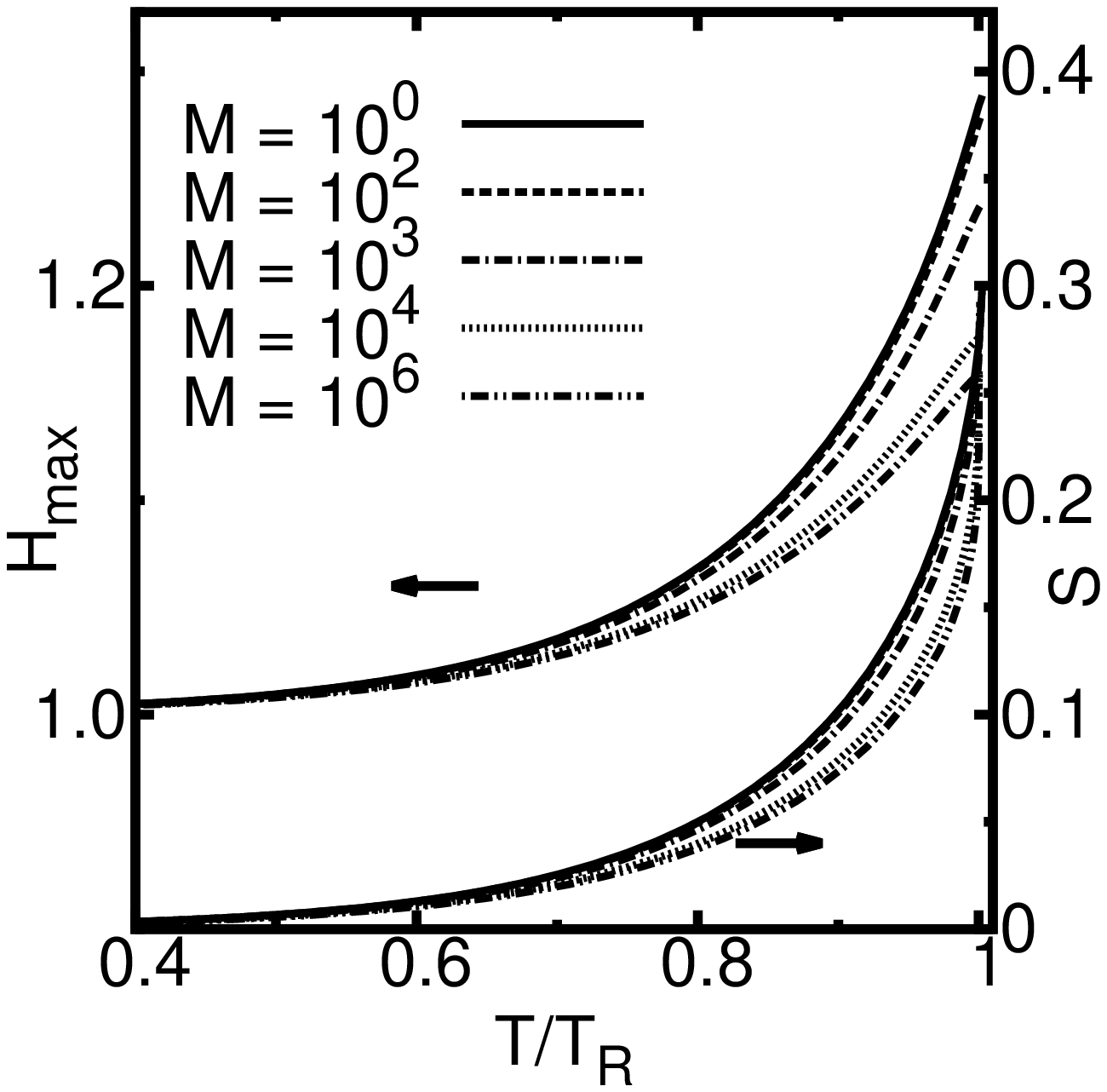}} \\
\end{tabular}
\caption{\label{fig:atrupture} 
      (A) Variation of dimensional viscosity $\mu$ with film thickness $h$ 
      (B) Morphologies of films at rupture with thickness $2$ nm and different film to base viscosity ratio $M$ in 
          a domain of size $\lambda_{m}$. 
      (C) Steep decrease in values of $H_{min}$ 
      (D) Flatter profiles with lower values of $H_{max}$ and $S$ are found for films with high values of $M$.}
\end{center}
\end{figure}

%%%%%%%%%%%%%%%%%%%%%%%%%%%%%%%%%%%%%
%\section{Mathematical Modeling}
A liquid film spontaneously dewets an underlying substrate under the influence of favorable excess intermolecular 
forces if it is thinner than their effective range ($\sim100$ nm). 
The dewetting process can be thought of as a combination of (i)~a pre-rupture stage or early stage which is before 
the formation of first dry spot or hole on the solid followed by (ii)~a post-rupture stage where the holes grow 
laterally and interacts with neighbouring growing holes to trap long cylindrical liquid ribbons in between.   
These ribbons break up to form droplets which become spherical under the influence of interfacial tension.
So the result of dewetting process is a collection of disjointed liquid droplets on the dewetted solid.
\cite{Reiter1992,Khanna1997}.
The simplest force field which allows to study dewetting is long range van der Waals attraction combined with much 
shorter range Born repulsion
\cite{Khanna1997}.
The excess intermolecular forces per unit area $\phi$, related to the excess free energy $\Delta G$ per unit area via 
($\phi=\frac{\partial\Delta G}{\partial h}$), is given by 
$\phi=-2S^{LW}(\frac{d_{0}^2}{h^3})\left[1-\left(\frac{l_{0}}{h}\right)^6\right]$ for this force field. 
Here $S^{LW}$ is the spreading coefficient of the film liquid on the substrate, $d_{0}$ is a cut-off distance 
($0.158$ nm) and $l_{0}$  is chosen such that  $\phi(l_{0}) = 0$  and  $\Delta G\left(l_{0}\right)=S^{LW}$.
We consider the thickness dependent viscosity, $\mu$, to be given by 
$\mu\left(h\right) = \mu_{b} + (\mu_{f} -\mu_{b})(1 - e^{-\left(\left(h-l_{0}\right)/b\right)^{\beta}})$ 
which is similar to the form reported in the experimental findings
\cite{Bodiguel2006}.  
Here $\mu_{f}$ is the bulk viscosity obtained as $h \rightarrow \infty$ and $\mu_{b}$ is the base viscosity or the 
viscosity at the substrate obtained as $h \rightarrow  l_{0}$, $b$ is related to the radius of gyration and the 
constant $\beta$ is taken as $6.8$ in the simulations. 
Some representative variations in viscosity for a given bulk viscosity, $\mu_{f}$, of $10^{9} Pa.s$ are shown in 
fig.~\ref{fig:atrupture}A.
Nodimensional viscosity, $\eta = \mu/\mu_{b}$ , is given by 
$\eta = 1 + (M - 1)(1 - e^{-((H-L)/B)^{\beta}})$. Where $H$, $L$ and $B$ are the nondimensional thicknesses with respect to mean thickness $h_{0}$ and
$M = \mu_{f}/\mu_{b}$ is the bulk to base viscosity ratio. 
Fig.~\ref{fig:atrupture}A shows that higher value of $M$ mean  reduced viscosity at the substrate ($\mu_{b}$) for the same bulk viscosity ($\mu_{f}$).
The evolution equation for the incompressible thin film with anisotropic viscosity effects obtained from Navier 
Stokes equations after long wave approximation and considering kinematic, shear free and capillary pressure boundary 
conditions at the free surface and no-slip and impermeability boundary conditions at the film-substrate interface is 
given by 
\begin{equation}
3h_{t}+\vec \nabla \cdot [h^{3}\vec \nabla (\gamma \nabla^{2}h-\vec \nabla \phi)/\mu(h)]=0
\label{Nav_dim}
\end{equation}
%%where $h(x,t)$ is the local thickness of the film, $\gamma$ is the surface tension of the film and $\phi$ is the 
%%excess intermolecular force as discussed above. 
%%

\begin{figure}[htbp]
\begin{center}
\begin{tabular}{cc}
\resizebox{2mm}{!}{A} &
\resizebox{2mm}{!}{B} \\
\resizebox{40mm}{!}{\includegraphics[width=1.5in, angle = -90]{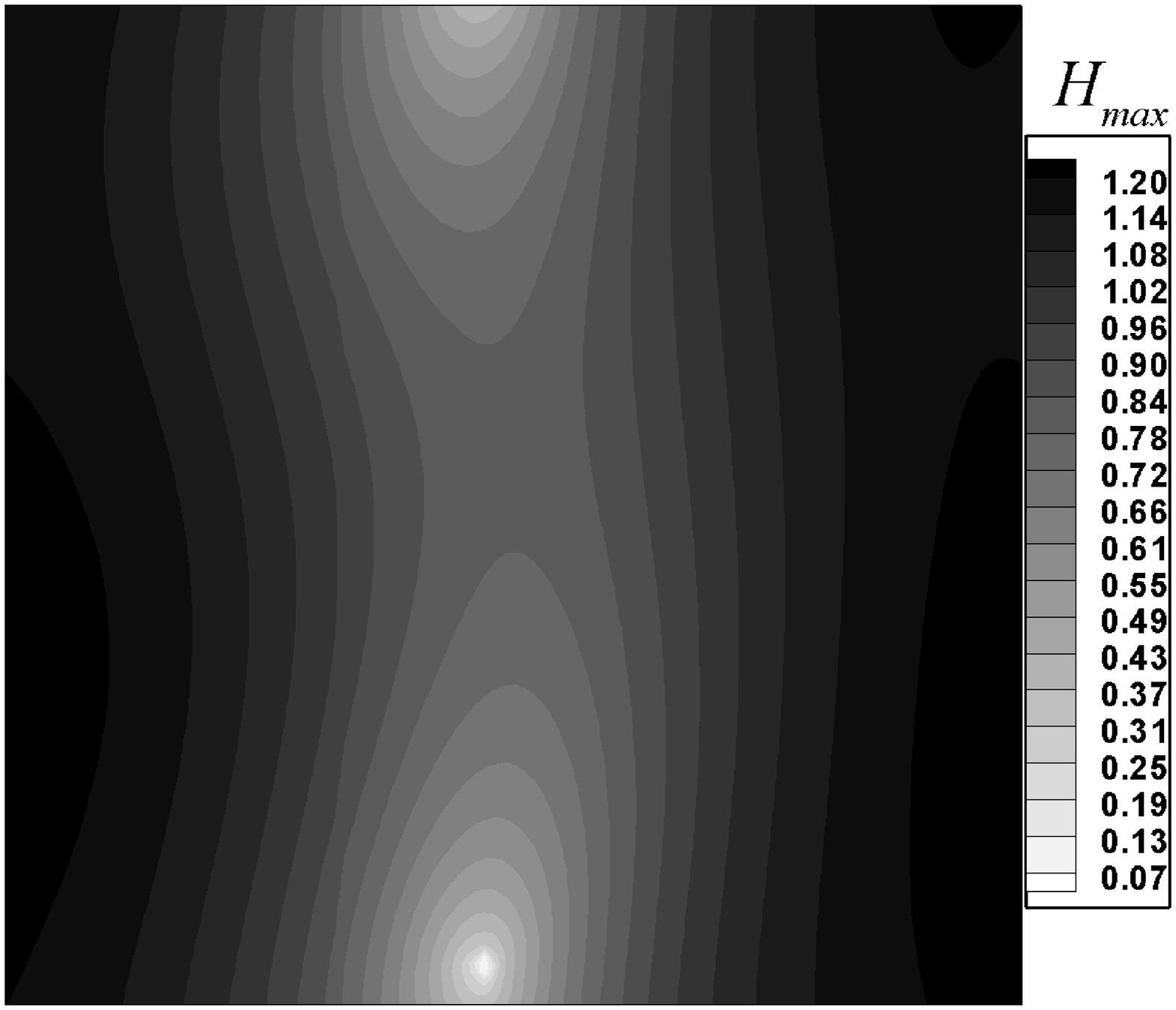}} &
\resizebox{40mm}{!}{\includegraphics[width=1.5in, angle = -90]{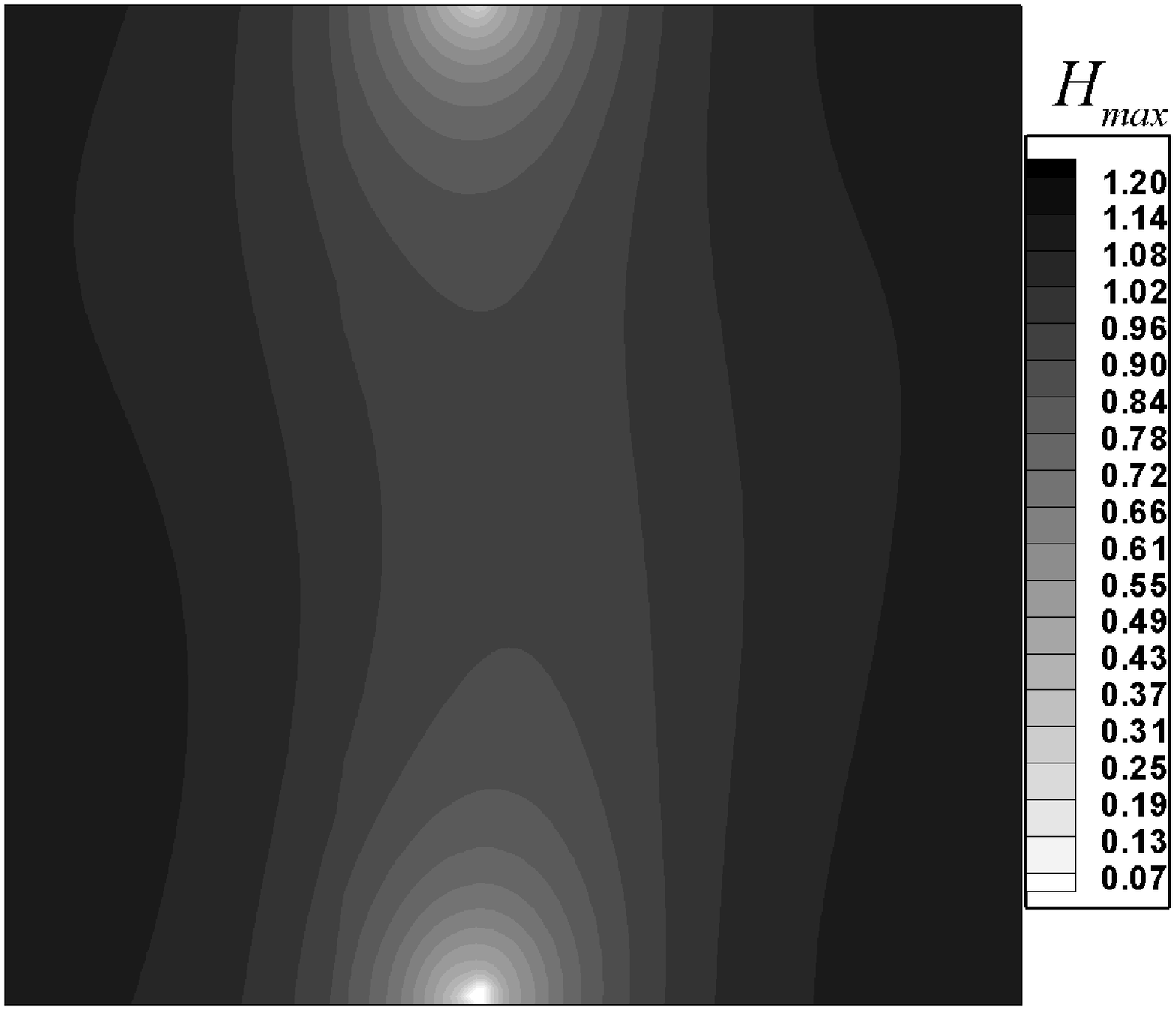}} \\
\end{tabular}
\caption{\label{fig:rup}
          3D morphological patterns at rupture for $2$ nm mean film thickness for
         (A) constant viscosity and 
         (B) high value of viscosity ratio ($M=10^6$) respectively.}
\end{center}
\end{figure}

The spinodal lengthscale or the dominant wavelength of the instability as given by linear stability analysis is the 
same as that of the constant viscosity films and is given as 
$\lambda_{m} = 2 \pi \left(-\frac{2 \gamma}{\phi_{h_{0}}}\right)^{1/2}$.
Here $\phi_{h_{0}}$ is the spinodal parameter (derivative of $\phi$) evaluated at $h = h_{0}$.
The linear time of rupture marked by trough of the spinodal wavelength reaching the substrate in linear analysis is 
given as $t_{r}={\frac{\gamma \mu(h_{0}) h_{0}^{5}}{3 (S^{LW} d_{0}^{2} \left[1-3 \left(\frac{l_{0}}{h_{0}}\right)^{6}\right])^{2}}}
ln(\frac{\mid h_{0}-l_{0} \mid}{\epsilon})$. It is clear that films with reduced viscosity e.g., films less than about $6$ nm in  
fig.~\ref{fig:atrupture}A are expected to show a correspondingly reduced linear time of rupture.
The nonlinear time of rupture will also be reduced as it is intimately related to the linear one
\cite{Khanna1997}. Simulations were performed on 2D and 3D versions of the nondimensionalized form of the evolution
equation%% equation so and so and include vector form of the nondimensional equation here.
\begin{equation}
H_{T}+\vec \nabla \cdot [H^{3}\vec \nabla (\nabla^{2}H-\vec \nabla \Phi)/\eta[H]]=0
\label{Nav_ndim}
\end{equation}
The non-dimensionalized parameters are given by
$H=\frac{h}{h_{0}}$, $\vec X=\frac{1}{h_{0}}(\frac{d_{0}}{h_{0}})(\frac{6 \mid S^{LW} \mid}{\gamma})^{1/2}\vec x$, 
$T=(\frac{d_{0}}{h_{0}})^4(\frac{12}{\mu_{b}\gamma h_{0}})(\mid S^{LW} \mid)^{2} t$ and 
$\Phi=\frac{1}{6 \mid S^{LW} \mid }(\frac{h_{0}^{3}}{d_{0}^{2}})\phi$
\cite{Khanna1997}.
Simulations were carried out in $n \times \lambda_{m}$ sized linear and square domains ($n = 0.5, 1,2,...$)
by finite differencing in space using successive half-node central differencing and integration in time using 
NAG library's subroutine D02EJF.
An ADI scheme was employed for 3D simulations.
A grid density of 192 points per $\lambda_{m}$ for $1$ nm thick film was found to be satisfactory.
Thicker films were simulated with a grid density varying as square of the film thickness appropos the 
nondimensionalization.

We first show that the early stages of dewetting  carries signature of slipping films even though one is simulating 
non-slipping films.
It is now established that the early stage of dewetting in slipping films are characterized by faster rupture, 
flatter morphologies around the growing depression and larger spinodal wavelengths
\cite{Gunter1994,Khanna1996,Gunter2000,Kajari2004,Karin2005,Karin2007}.
Fig.~\ref{fig:atrupture}B illustrates the surface morhpology at the time of rupture of a $2$ nm film as
the viscosity ratios $M$ increases from unity (constant viscosity case) to a high value of $10^{6}$. 
The film around the hole becomes progressively flatter as one increases the $M$ by reducing the $\mu_{b}$ while keeping the same $\mu_{f}$.
This is due to the dispersal of lesser material into the surrounding film as the reduced viscosity at thinner 
regions dictates that lesser liquid needs to be scooped to form the hole. 
This is evidenced by the increasing sharpness of the hole profile as $M$ increases.
Also, faster hole formation as shown in fig.~\ref{fig:atrupture}C means that the crest of the spinodal wave do not get a chance 
to develop fully before the hole forms.
The evolution of surface morphology as marked by maximum non dimensional thickness, $H_{max}$, and root mean 
square surface roughness, $S$, also corroborate this (refer to Fig.~\ref{fig:atrupture}D).
Both evolve to a lesser extent indicating flatter morphology as $M$ is increased.
The spinodal wavelength here remains unchanged from that of constant viscosity film.
In slipping films the spinodal wave length increases considerably
\cite{Kajari2004} 
and this fact can be exploited to differentiate between slipping films and films with varaiable viscosity.
However, as the spinodal wavelength itself is a crucial input to the estimate of the often unknown excess 
intermolecular force field, one can not really know if it has increased or reduced.
Same is the case with reduced time scales also
\cite{Sharma2003}.
The flatness of the surrounding film which remains free of these issues and can serve as a differentiating
marker remains the same in both cases.  
The flatness of the surrounding film, though, is not easy to resolve in experiments.
Even in simulations, it is not easy as shown in fig.~\ref{fig:rup} which presents the contour 
map of film's surface at the time of hole formation for constant viscosity film and a film having very high value of $M$. 
If one observes closely, one can see more contours in between the holes for Fig.~\ref{fig:rup}A indicating different depths of the film surface.
 Where as for higher $M$ (Fig.~\ref{fig:rup}B) the lesser number of conours in between the depression demonstrate a sharper hole profiles with 
flatter surroundings.\\
%%%
\begin{figure}[htbp]
\begin{center}
\begin{tabular}{cc}
\resizebox{2mm}{!}{A} &
\resizebox{2mm}{!}{B} \\
\resizebox{40mm}{!}{\includegraphics[width=1.5in]{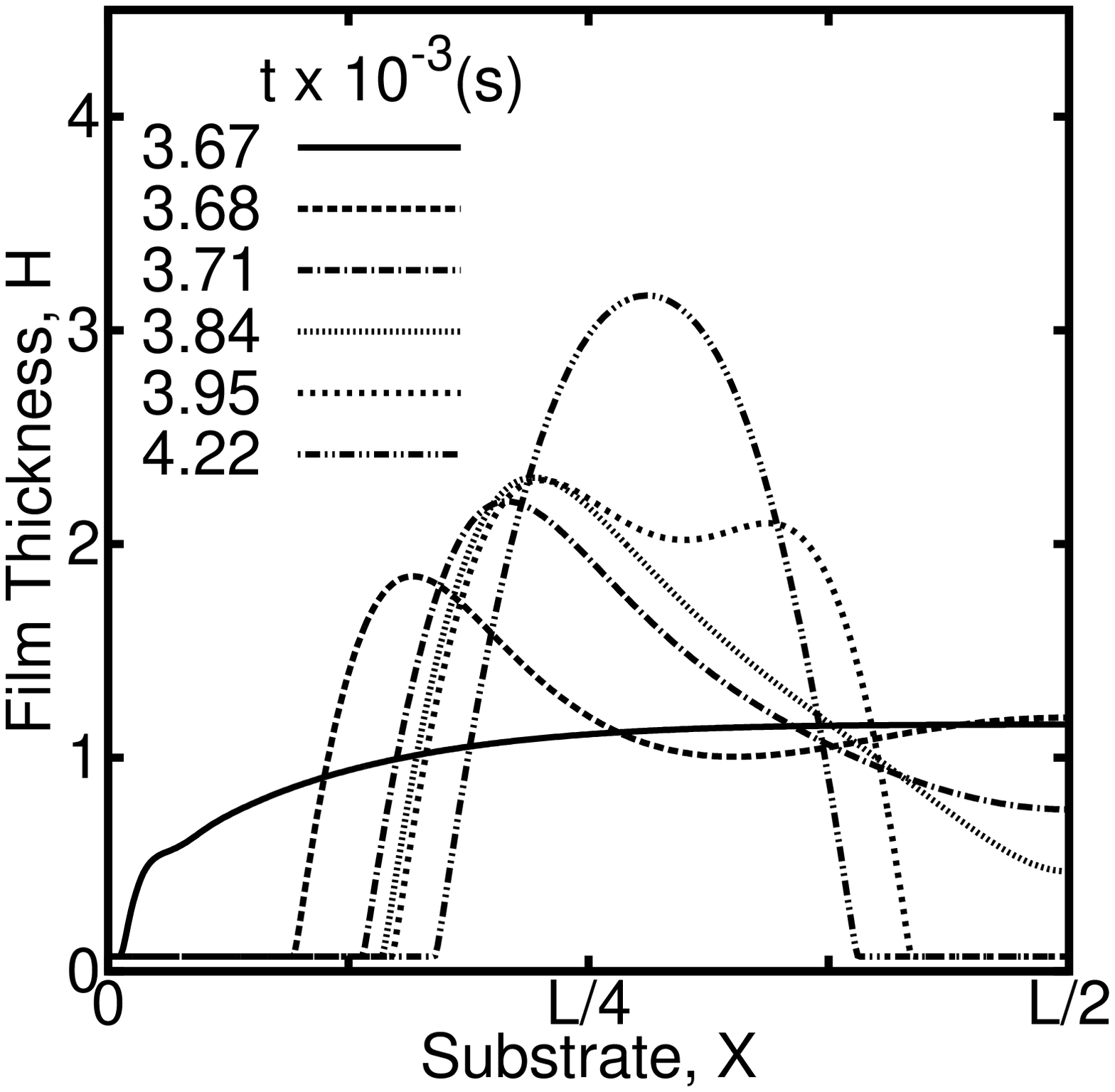}} &
\resizebox{40mm}{!}{\includegraphics[width=1.5in]{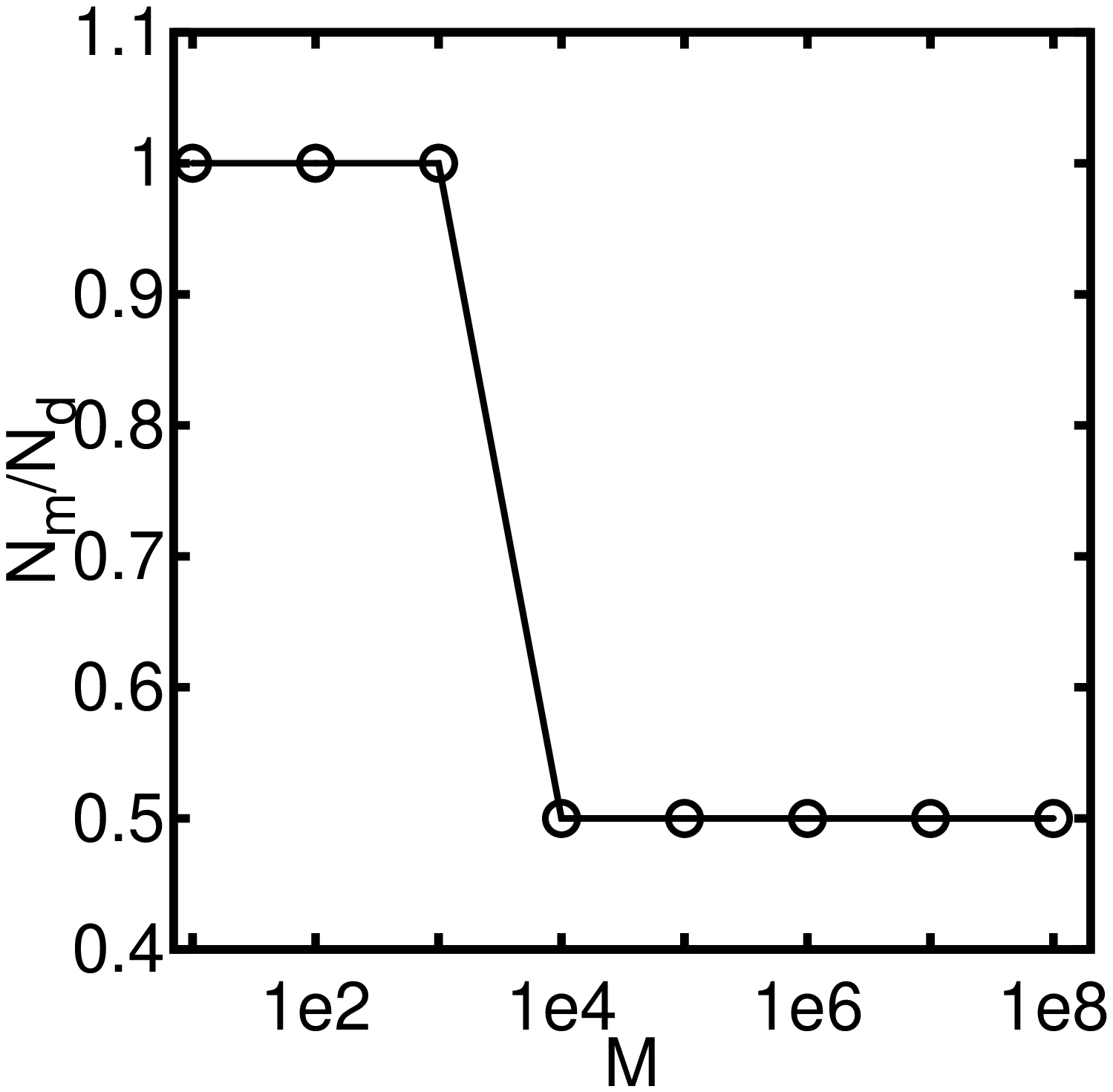}} \\
\end{tabular}
\caption{\label{fig:evolv} Hole growth phase after rupture. 
         (A) Hole growth for a film of thicnkess $2$ nm simulated in a domain of size $\lambda_{m}/2$ with 
             viscosity ratio $M=10^{6}$. Kinetics of the film helps in formation of satellite hole.
         (B) Effect of $M$ on the ratio of lengthscales formed at equilibrium to that predicted by linear stability 
             analysis for films of thickness $2$ nm in a domain of size $2 \lambda_{m}$.}
\end{center}
\end{figure}

\begin{figure}[htbp]
\begin{center}
\includegraphics*[width=1.7in]{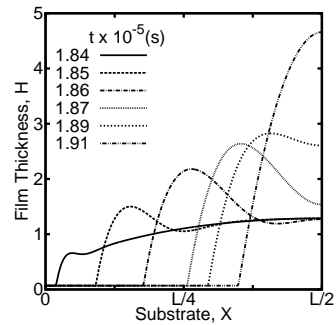} 
\end{center}
\vspace*{-12pt}
\caption{\label{fig:evolv1} Hole growth phase after rupture. 
           Hole growth for a film of thicnkess $2$ nm simulated in a domain of size 
             $\lambda_{m}/2$ with viscosity ratio $M=10$.}
\end{figure}
We now show the emergence of sub-spinodal lengthscale in the later stages of dewetting.
Fig.~\ref{fig:evolv} presents the results for post-rupture variations in morphology of a $2$ nm film with a very high 
viscosity ratio of $10^{6}$.
The simulations are done in one-half of the spinodal wavelength for faster simulations as well as better
spatial resolution. 
Symmetry boundary conditions at both ends were used to have zero liquid flux at the edges. 
These were preferred to usual periodic boundary conditions which are relevant whenever simulations are done in 
integral multiples of spinodal wavelengths.
\begin{figure}
\begin{center}
\begin{tabular}{cc}
\resizebox{2mm}{!}{A}& 
\resizebox{2mm}{!}{B} \\
\resizebox{30mm}{!}{\includegraphics[width=1.4in, angle = -90]{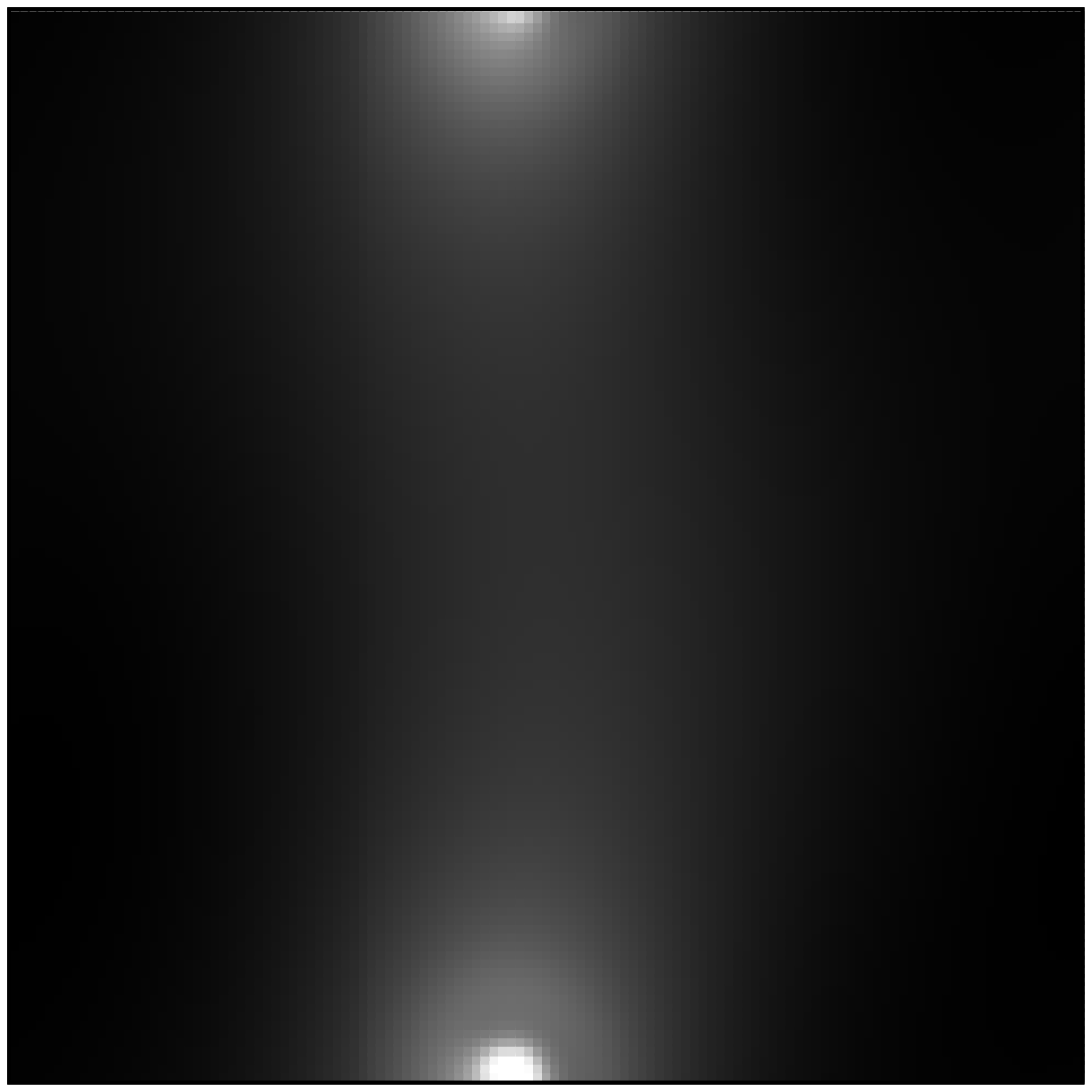}}& 
\resizebox{30mm}{!}{\includegraphics[width=1.4in, angle = -90]{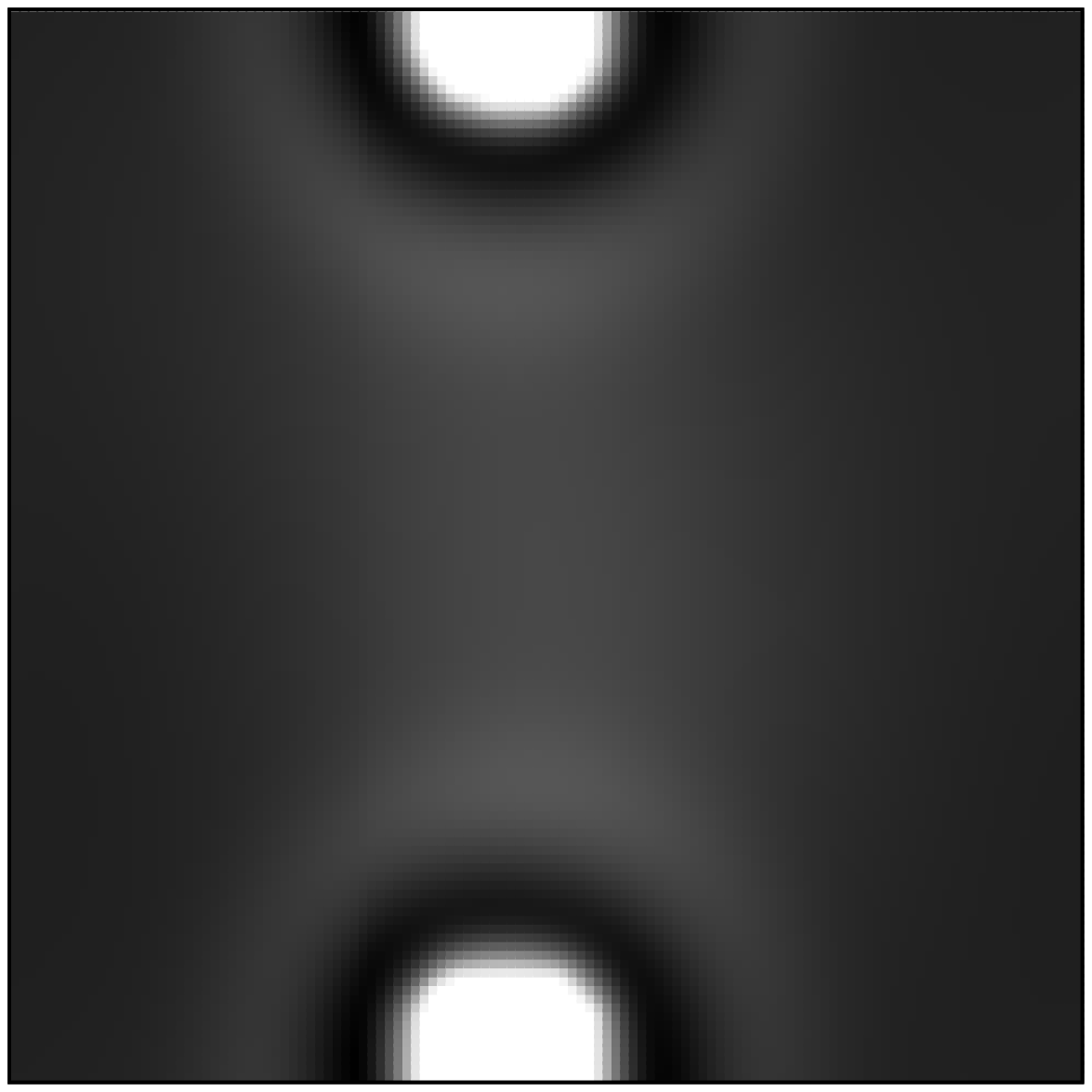}} \\
  & \\
\resizebox{2mm}{!}{C} &
\resizebox{2mm}{!}{D} \\
\resizebox{30mm}{!}{\includegraphics[width=1.4in, angle = -90]{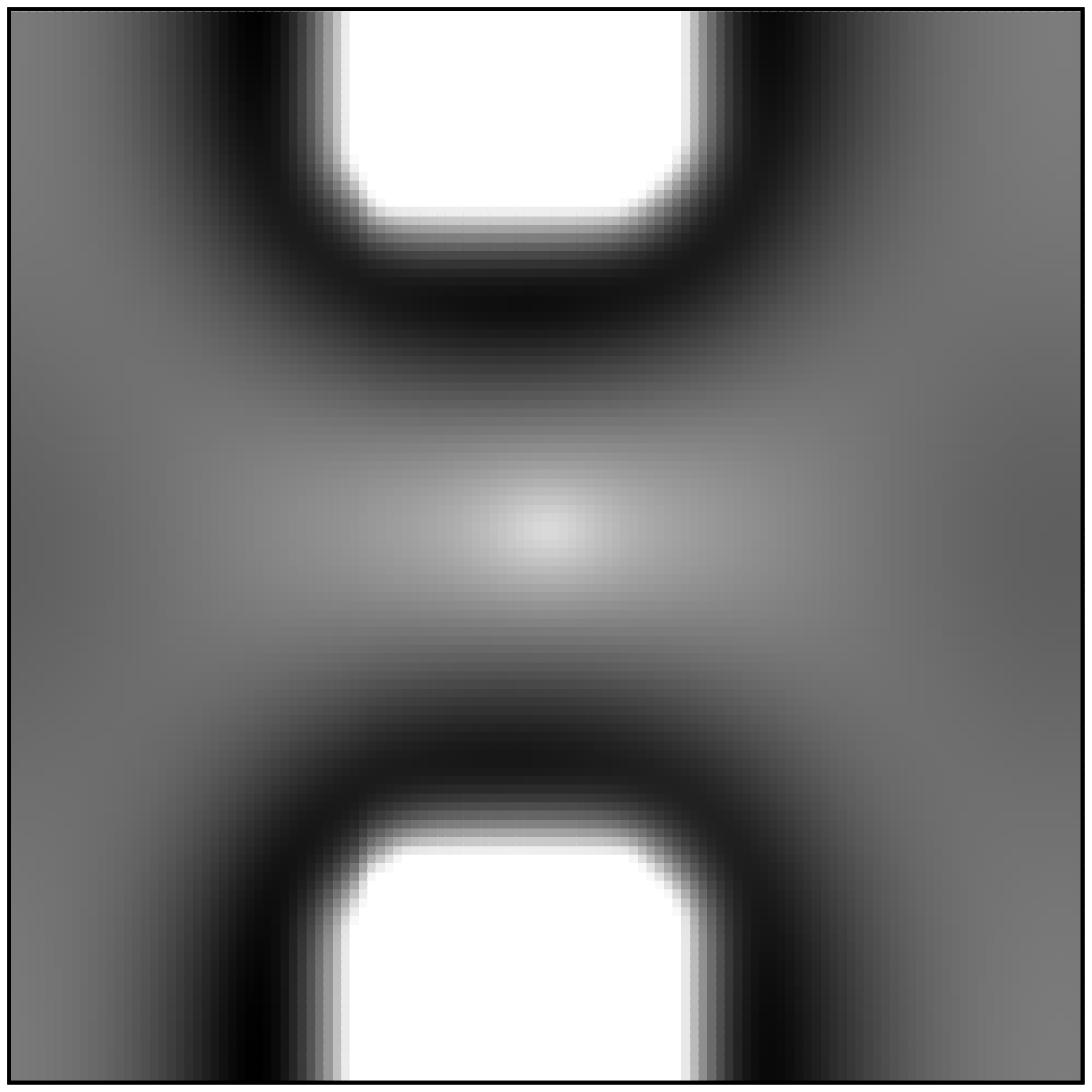}} &
\resizebox{30mm}{!}{\includegraphics[width=1.4in, angle = -90]{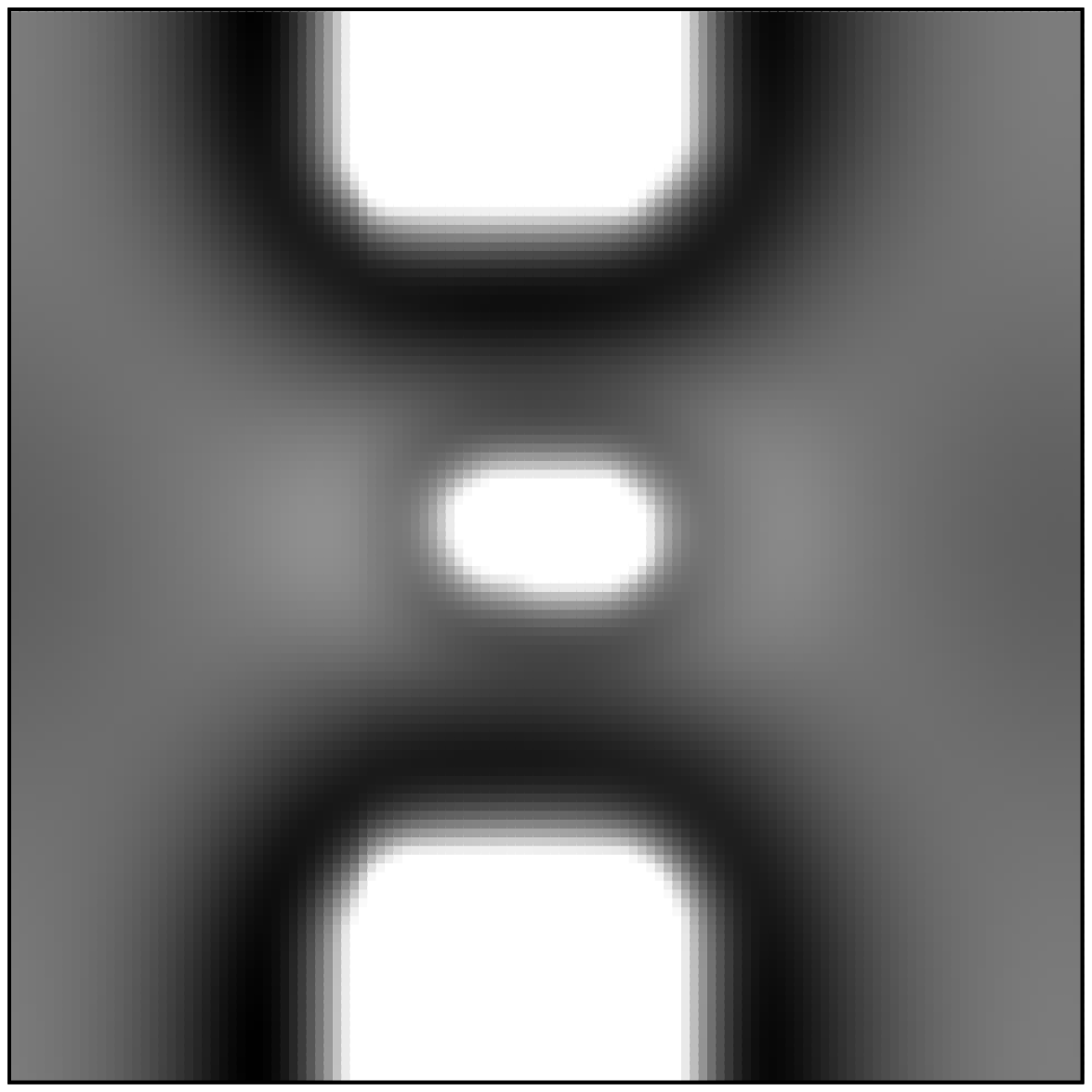}} \\
\end{tabular}
\caption{\label{fig:tp2evolv} 
         3D morphological evolutions for film of thickness $2$ nm and viscosity ratio $M=10^6$ undergoing 
         satellite hole formation. 
         The dimensional time associated with the different stages of evolution are 
         A)1012 s, B)1012.73 s, C)1024.49 s, D)1025.88 s respectively.}
\end{center}
\end{figure}
The hole grows as reported extensively in the past
\cite{Reiter1992,Khanna1997}  
and mismatch between the rate of dewetting and subsequent dispersal on the surrounding film forms an
asymmetric elevated rim at the hole edge with a depression on the far side ($t = 3.68\times 10^{3}$s).
The further part of the film shows no effect.
Soon, the depression reaches the end of the domain and the growing hole starts interacting with the neighbouring 
hole (not shown).
At this point the depression deepens further to form a satellite hole ($t=3.84\times 10^{3}$s, $3.95\times 10^{3}$s, $4.22\times 10^{3}$s) which 
also starts growing.
The liquid trapped between the satellite hole and the primary hole becomes increasingingly circular under the 
influence of interfacial tension and forms a droplet.
Thus, one full droplet emerges in only one-half of the spinodal wavelength making two droplets per spinodal 
wavelength. 
Large domain simulations with periodic boundary conditions (morphologies not shown here) also confirm the formation of 
satellite holes and subsequent increase in the number of droplets per spinodal wavelength. 
In complete contrast to this, films with lesser viscosity ratio follow the normal dewetting process. Both these effects can be seen from 
Fig.~\ref{fig:evolv}B which shows that for smaller $M$ values the number of holes formed ($N_{d}$) is the same as predicted by linear stability 
analysis($N_{m}$). However, the ratio $N_{m}/N_{d}$ is much smaller for higher value of $M$ indicating sub-spinodal lenghtscales due to 
formation of satellite holes.
Fig.~\ref{fig:evolv1} presents results for $M = 10$ where the depression at the far end of the rim does not deepen
to form a satellite hole but rises up to form a liquid ribbon between two growing primary holes ($t=1.91\times 10^{5}$s).
In this case the depression gets filled up due to pouring of the liquid from the rims of the growing
primary neighbouring holes before it can thin under the influence of excess intermolecular forces.
This influx of liquid into the intervening depression is not possible for high values of $M$ due to higher 
visocosity at the rims and lower viscosity at the depression.
This combined with the ability of forming sharper holes (Fig.~\ref{fig:atrupture}) by removal of lesser material manifests 
into formation of satellite holes for these films. 
\begin{figure}
\begin{center}
\includegraphics*[width=1.7in]{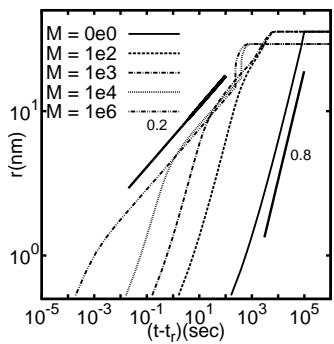} 
\end{center}
\vspace*{-12pt}
\caption{\label{fig:rad} 
           Dewetted length ($r$) at different times after ruprture for a film of thickness $2$ nm simulated in a domain of size 
             $\lambda_{m}/2$ for different viscosity ratios $M$. The bold lines are the slopes and hence indicative of the growth law
             exponents ($\alpha$) for different $M$.}
\end{figure}
Formation of satellite holes is confirmed by 3D simulations also.
Simulations were carried out in a unit cell of spindal wavelength side with periodic boundary conditions for the same $2$ nm 
thick film. Fig.~\ref{fig:tp2evolv} which shows grayscale map of the surface morphology of the film shows deepening of the intervening 
depression into a satellite hole between two growing primary holes.

The rate of dewetting can be given by growth law $r \propto t^{\alpha}$ where $r$ is the dewetted length.
Fig.~\ref{fig:rad} represents the growth of dewetted length in $2$D simulations which has contributions from both primary and
satellite holes (if and when formed).
For constant viscosity case it is well known that the exponent $\alpha$ has much higher values of $0.8$ and above 
\cite{Khanna1997,Ghatak1999} as also seen from Fig.~\ref{fig:rad}.
 However, films with higher $M$ the exponent has a much lower value ($\sim 0.2$) for $M=10^{6}$.
Thus, even the growth law shares the tendancy of slipping films to have lower exponents. 
However, the exponents for slipping films decrease to about $2/3$ only
\cite{Gunter2013}. 
Previous studies of dewetting of non-Newtonian thin films with shear thinning or shear thickening effects have shown to
display only change in rupture times without any change in lengthscales
\cite{Sharma2003}. 
For the first time it is shown in this letter that viscosity decreasing with decreasing film thickness leads to nonlinear 
mobility factor which now has a noncubic variation with thickness and has important effects on the thin film instability.
Based on the simulations it is clear that these non-slipping, viscosity varying films on homogeneous substrate show 
sharper holes, flatter morphologies, faster rupture at pre-rupture stage like slipping films.
Post rupture these films lead to formation of satellite holes giving rise to sub-spinodal length scales which so far was considered 
to be only possible for films dewetting on heterogeneous substrates or undergoing nucleation.
Thus, if one uses the dewetting of thin films as a cantilever to understand the underlying force field in case of 
films with anisotropic viscous effects is bound to miscalculate the force field and would consider it to be much stronger 
than it really is if the graded viscosity effect is neglected. 
On the other hand this retarded viscosity effect can be beneficially utilized to fabricate miniaturized patterns.

%%%%%%%%%%%%%%%%%%%%%%%%%%%%%%%%%%%%%%%%%%%%%%%%%%%%%
%\section{Conclusions}
%Previous studies of dewetting of thin films with viscousity dependant thickness effects have shown to display only change in 
%time of rupture without any change in lengthscale. 
%For the first time it is shown in this letter that viscosity decreasing with decreasing film thickness effects at the 
%film substrate interface leads to nonlinear mobility factor which now has a noncubic variation with thickness. 
%Anisotropic viscosity effects are found to reduce the rupture time and yields less rough surface profile at rupture 
%show signature of slipping films and further leads to satellite hole formation can be observed as in nucleation assisted 
%rupture or films in defect sensitive spinodal regions for heterogeneous substrates, though we can see that the real 
%reason is none of the above. 
%Thus, if one uses the dewetting of thin films as a cantilever to understand the underlying force field in case of 
%anisotropic films is bound to miscalculate the force field and would consider it to be much stronger (about $4$ to 
%$9$ times more stronger) than it really is if the graded viscosity effect is neglected. 
%On the other hand this retarded viscosity effect can be beneficially utilized to miniaturize patterns in dewetting 
%liquid thin films. 

\bibliography{thickdepvisc}
\end{document}